\documentclass[twoside,11pt]{3ssmmp} 

\usepackage{fancyhdr}    
\usepackage{graphicx}
\usepackage{rotating}

\renewcommand{\headrulewidth}{0.4pt}

\begin{document}
 \baselineskip=11pt

\title{Renormalizable Theories from Fuzzy Higher Dimensions
\hspace{.25mm}\thanks{\,Partially supported by the European
Commission under the RTN contract MRTN-CT-2004-503369, by the
programmes Irakleitos and Pythagoras of the Greek Ministry of
Education and the NTUA programme Protagoras.}}
\author{\bf{Paolo Aschieri}\hspace{.25mm}\thanks{\,e-mail address:
aschieri@theorie.physik.uni-muenchen.de}
\\ \normalsize{Dipartimento di Scienze e Tecnologie
Avanzate,}\\\normalsize{Universit{\'a} del Piemonte Orientale, and INFN,}\\
\normalsize{Corso Borsalino 54, I-15100,  Alessandria, Italy}
\\\normalsize{Max-Planck-Institut f\"{u}r Physik }
\\\normalsize{F\"{o}hringer Ring 6, D-80805 M\"{u}nchen}\vspace{2mm}\\
\bf{John Madore}\hspace{.25mm}\thanks{\,e-mail address:
John.Madore@th.u-psud.fr}
\\ \normalsize{Laboratoire de Physique
Th\'{e}orique }\\\normalsize{Universit\'{e} de Paris-Sud,}\\
\normalsize{B\^{a}timent 211, F-91405 Orsay } \vspace{2mm}
\\ \bf{Pantelis Manousselis}\hspace{.25mm}\thanks{\,e-mail
address: pantelis@upatras.gr}
\\ \normalsize{Department of Engineering Sciences}\\\normalsize{ University of Patras, 26110 Patras} \vspace{2mm} \\
\bf{George Zoupanos}\hspace{.25mm}\thanks{\, On leave from Physics
Department, National Technical University Zografou Campus,
GR-15780 Athens, e-mail address: zoupanos@mail.cern.ch}
\\\normalsize{Max-Planck-Institut f\"{u}r Physik } \\\normalsize{F\"{o}hringer Ring 6,}\\
\normalsize{D-80805 M\"{u}nchen}\\ \normalsize{Sektion Physik,}
\\\normalsize{Universit\"{a}t M\"{u}nchen }\\\normalsize{Theresienstra{\ss}e 37, D-80333
M\"{u}nchen } }

\date{}

\maketitle
\begin{abstract}
We consider gauge theories defined in higher dimensions where the
extra dimensions form a fuzzy space (a finite matrix manifold). We
reinterpret these gauge theories as four-dimensional theories with
Kaluza-Klein modes.  We then perform a generalized \`a la
Forgacs-Manton dimensional reduction. We emphasize some striking
features emerging such as (i) the appearance of non-abelian gauge
theories in four dimensions starting from an abelian gauge theory
in higher dimensions, (ii) the fact that the spontaneous symmetry
breaking of the theory takes place entirely in the extra
dimensions and (iii) the renormalizability of the theory both in
higher as well as in four dimensions.
\end{abstract}
\section{Introduction}
In the recent years a huge theoretical effort has been devoted
aiming to establish a unified description of all interactions
including gravity. Out of this sustained endeavor, along the
the superstring theory framework
\cite{Green:1987sp}, the non-commutative geometry one has emerged \cite{Connes,
Madore}. An interesting development worth noting was the
observation that a natural realization of non-commutativity of
space appears in the string theory context of D-branes in the
presence of a constant antisymmetric field \cite{Seiberg:1999vs},
which brought together the two approaches. However these very
interesting approaches do not address as yet the usual problem of
the Standard Model of Elementary Particle Physics, i.e. the
presence of a plethora of free parameters to the ad-hoc
introduction of the Higgs and Yukawa sectors in the theory. These
sectors might have their origin in a higher-dimensional theory
according to various schemes among whose the first one was the
Coset Space Dimensional Reduction (CSDR) \cite{Forgacs:1979zs,
Kapetanakis:hf, Kubyshin:vd}. The CSDR scheme has been used to
reduce in four dimensions a ten-dimensional, $N=1, E_8$ gauge
theory \cite{Kapetanakis:hf, Manousselis:2001re,
Manousselis:2004xd}and might be an appropriate reduction scheme of
strings over nearly Kaehler manifolds \cite{Micu:2004tz}. More
recently  the dimensional reduction of gauge
theories defined in higher dimensions where the extra dimensions form a
fuzzy coset (a finite matrix manifold) has been examined \cite{Aschieri:2003vy,
Aschieri:2004vh}. This might lead to interesting
constructions for Particle Physics models .
In the present paper we would like to present
the main points and results of the dimensional reduction over
fuzzy coset spaces and emphasize in particular the issue of the
renormalizability of gauge theories defined with such a non-commutative setup.

In the  CSDR one assumes that the form of space-time is
$M^{D}=M^{4} \times S/R$ with $S/R$ a homogeneous space (obtained
as the quotient of the Lie group $S$ via the Lie subgroup $R$).
Then a gauge theory with gauge group $G$ defined on $M^{D}$ can be
dimensionally reduced to $M^{4}$ in an elegant way using the
symmetries of $S/R$, in particular the resulting four-dimensional
gauge group is a subgroup of $G$. Although the reduced theory in
four dimensions is power counting renormalizable the full
higher-dimensional theory is non-renormalizable with dimensionful
coupling. The CSDR scheme reduces dimensionally a gauge theory
with gauge group $G$ defined on $M_4 \times S/R$ to a gauge theory
on $M_4$ imposing the principle that fields should be invariant
under the $S$ action up to a $G$ gauge transformation. The CSDR
scheme constitutes an elegant and consistent truncation of the
full theory in four dimensions, keeping  only the first terms of
the field expansion in higher harmonics of the  compact coset
spaces. When keeping all the higher harmonics, i.e. the
Kaluza-Klein modes in four dimensions, then the theory in general
becomes non-renormalizable as expected, since the theory was
originally defined in higher than four dimensions. Still it is
very interesting the fact that one can discuss the dependence of
the couplings of the theory on the cutoff, or the beta-function of
the couplings in the Wilson renormalization scheme
\cite{Dienes:1998vg, Kubo:1999ua, Kubo:2000hy}. In the fuzzy-CSDR
we apply the CSDR principle in the case that the extra dimensions
are a finite approximation of the homogeneous spaces $S/R$,  i.e.
a fuzzy coset. Fuzzy spaces are obtained by deforming the algebra
of functions on their commutative parent spaces. The algebra of
functions (from the fuzzy space to complex numbers) becomes finite
dimensional and non-commutative, indeed it becomes a matrix
algebra. Therefore, instead of considering the algebra of
functions $Fun(M^{D})\sim Fun(M^{4}) \otimes Fun(S/R)$ we consider
the algebra $A= Fun(M^{4}) \otimes Fun((S/R)_F)$ where
$Fun(M^{4})$ is the usual commutative algebra of functions on
Minkowski space $M^{4}$ and $Fun((S/R)_F)=M_N$ is the finite
dimensional non-commutative algebra of $N\times N$ matrices that
approximates the functions on the coset $S/F$. On this finite
dimensional algebra we still have the action of the symmetry group
$S$; this very property allows us to apply the CSDR scheme to
fuzzy cosets. The reduction of a gauge theory defined on $M^4
\times (S/R)_F$ to a gauge theory on $M^4$ is a two step process.
One first rewrites the higher-dimensional fields, that initially
depends on the commutative coordinates $x$ and the noncommutative
ones $X$, in terms of only the commutative coordinates $x$,  with
the fields now being also $N\times N$ matrix valued. One then
imposes the fuzzy-CSDR constraints on this four-dimensional
theory. We can say that the theory is a higher-dimensional theory
because the fuzzy space $(S/R)_F$ is a noncommutative
approximation of the coset space $S/R$; in particular the spatial
symmetry group $S$ of the space $(S/R)_F$ is the same as that of
the commutative space $S/R$. However the noncommutative theory has
the advantage of being power counting renormalizable because
$Fun((S/R)_F)$ is a finite dimensional space; it follows that also
after applying the fuzzy-CSDR scheme we obtain a power counting
renormalizable theory.

\section{The Fuzzy sphere}
The fuzzy sphere \cite{Mad, Madore} is a matrix approximation of
the usual sphere $S^2$. The algebra of functions on $S^2$ (for
example spanned by the spherical harmonics) is truncated at a
given frequency and thus becomes finite dimensional. The
truncation has to be consistent with the associativity of the
algebra and this can be nicely achieved relaxing the commutativity
property of the algebra. The fuzzy sphere is the ``space''
described by this non-commutative algebra. The algebra itself is
that of $N\times N$ matrices. More precisely, the algebra of
functions on the ordinary sphere can be generated by the
coordinates of {\bf{R}}$^3$ modulo the relation $
\sum_{\hat{a}=1}^{3} {x}_{\hat{a}}{x}_{\hat{a}} =r^{2}$. The fuzzy
sphere $S^2_{F}$ at fuzziness level $N-1$ is the non-commutative
manifold whose coordinate functions $i {X}_{\hat{a}}$ are $N
\times N$ hermitian matrices proportional to the generators of the
$N$-dimensional representation of $SU(2)$. They satisfy the
condition $ \sum_{\hat{a}=1}^{3} X_{\hat{a}} X_{\hat{a}} = \alpha
r^{2}$ and the commutation relations
\begin{equation}
[ X_{\hat{a}}, X_{\hat{b}} ] = C_{\hat{a} \hat{b} \hat{c}}
X_{\hat{c}}~,
\end{equation}
where $C_{\hat{a} \hat{b} \hat{c}}= \varepsilon_{\hat{a} \hat{b}
\hat{c}}/r$ while the proportionality factor $\alpha$ goes as
$N^2$ for $N$ large. Indeed it can be proven that for
$N\rightarrow \infty$ one obtains the usual commutative sphere.

On the fuzzy sphere there is a natural $SU(2)$ covariant
differential calculus. This calculus is three-dimensional and the
derivations $e_{\hat{a}}$ along $X_{\hat{a}}$ of a function $ f$
are given by $e_{\hat{a}}({f})=[X_{\hat{a}},
{f}]\,.\label{derivations}$ Accordingly the action of the Lie
derivatives on functions is given by
\begin{equation}\label{LDA}
{\cal L}_{\hat{a}} f = [{X}_{\hat{a}},f ]~;
\end{equation}
these Lie derivatives satisfy the Leibniz rule and the $SU(2)$ Lie
algebra relation
\begin{equation}\label{LDCR}
[ {\cal L}_{\hat{a}}, {\cal L}_{\hat{b}} ] = C_{\hat{a} \hat{b}
\hat{c}} {\cal L}_{\hat{c}}.
\end{equation}
In the $N \rightarrow \infty$ limit the derivations $e_{\hat{a}}$
become $
e_{\hat{a}} = C_{\hat{a} \hat{b} \hat{c}} x^{\hat{b}}
\partial^{\hat{c}}\,
$ and only in this commutative limit the tangent space becomes
two-dimensional. The exterior derivative is given by
\begin{equation}
d f = [X_{\hat{a}},f]\theta^{\hat{a}}
\end{equation}
with $\theta^{\hat{a}}$ the one-forms dual to the vector fields
$e_{\hat{a}}$,
$<e_{\hat{a}},\theta^{\hat{b}}>=\delta_{\hat{a}}^{\hat{b}}$. The
space of one-forms is generated by the $\theta^{\hat{a}}$'s in the
sense that for any one-form $\omega=\sum_i f_i d h_i \:t_i$ we
can always write
$\omega=\sum_{\hat{a}=1}^3{\omega}_{\hat{a}}\theta^{\hat{a}}$ with
given functions $\omega_{\hat{a}}$ depending on the functions
$f_i$, $h_i$ and $t_i$. The action of the Lie derivatives
${\cal  L}_{\hat{a}}$ on the one-forms $\theta^{\hat{b}}$ explicitly
reads
\begin{equation}\label{2.16}
{\cal L}_{\hat{a}}(\theta^{\hat{b}}) =  C_{\hat{a}\hat{b}\hat{c}}
\theta^{\hat{c}}~.
\end{equation}
On a general one-form
$\omega=\omega_{\hat{a}}\theta^{\hat{a}}$ we have $ {\cal
L}_{\hat{b}}\omega={\cal
L}_{\hat{b}}(\omega_{\hat{a}}\theta^{\hat{a}})=
\left[X_{\hat{b}},\omega_{\hat{a}}\right]\theta^{\hat{a}}-\omega_{\hat{a}}C^{\hat{a}}_{\
\hat{b} \hat{c}}\theta^{\hat{c}} $ and therefore
\begin{equation}
({\cal
L}_{\hat{b}}\omega)_{\hat{a}}=\left[X_{\hat{b}},\omega_{\hat{a}}\right]-
\omega_{\hat{c}}C^{\hat{c}}_{\ \hat{b}  \hat{a}}~;\label{fund}
\end{equation}
this formula will be fundamental for formulating the CSDR
principle on fuzzy cosets.

The differential geometry on  the product space Minkowski times
fuzzy sphere, $M^{4} \times S^2_{F}$, is easily obtained from that
on $M^4$ and on $S^2_F$. For example a one-form $A$ defined on
$M^{4} \times S^2_{F}$ is written as
\begin{equation}\label{oneform}
A= A_{\mu} dx^{\mu} + A_{\hat{a}} \theta^{\hat{a}}
\end{equation}
with $A_{\mu} =A_{\mu}(x^{\mu}, X_{\hat{a}} )$ and $A_{\hat{a}}
=A_{\hat{a}}(x^{\mu}, X_{\hat{a}} )$.

One can also introduce spinors on the fuzzy sphere and study the
Lie derivative on these spinors. Although here we have sketched
the differential geometry on the fuzzy sphere,  one can study
other (higher-dimensional) fuzzy spaces (e.g. fuzzy $CP^M$) and
with similar techniques their differential geometry.

\section{Actions in higher dimensions seen as four-dimensional actions
(Expansion in Kaluza-Klein modes)} First we consider on  $M^{4}
\times (S/R)_{F}$ a non-commutative gauge theory with gauge group
$G=U(P)$ and examine its four-dimensional interpretation.
$(S/R)_{F}$ is a fuzzy coset, for example the fuzzy sphere
$S^{2}_{F}$. The action is
\begin{equation}\label{formula8}
{\cal A}_{YM}={1\over 4g^{2}} \int d^{4}x\, kTr\, tr_{G}\,
F_{MN}F^{MN},
\end{equation}
where $kTr$ denotes integration over the fuzzy coset $(S/R)_F\,$
described by $N\times N$ matrices; here the parameter $k$ is
related to the size of the fuzzy coset space. For example for the
fuzzy sphere we have $r^{2} = \sqrt{N^{2}-1}\pi k$ \cite{Madore}.
In the $N\rightarrow \infty$ limit $kTr$ becomes the usual
integral on the coset space. For finite $N$, $Tr$ is a good integral
because it has the cyclic property $Tr(f_1\ldots f_{p-1}f_p)=Tr(f_pf_1\ldots
f_{p-1})$. It is also invariant under the action of the group $S$,
that is  infinitesimally given by the Lie derivative.
In the action (\ref{formula8}) $tr_G$ is the gauge group $G$ trace. The
higher-dimensional field strength $F_{MN}$, decomposed in
four-dimensional space-time and extra-dimensional components, reads
as follows $(F_{\mu \nu}, F_{\mu \hat{b}}, F_{\hat{a} \hat{b}
})\,;$ explicitly the various components of the field strength are
given by
\begin{eqnarray}
F_{\mu \nu} &=&
\partial_{\mu}A_{\nu} -
\partial_{\nu}A_{\mu} + [A_{\mu}, A_{\nu}],\\[.3 em]
F_{\mu \hat{a}} &=&
\partial_{\mu}A_{\hat{a}} - [X_{\hat{a}}, A_{\mu}] + [A_{\mu},
A_{\hat{a}}], \nonumber\\[.3 em]
F_{\hat{a} \hat{b}} &=&   [ X_{\hat{a}}, A_{\hat{b}}] - [
X_{\hat{b}}, A_{\hat{a}} ] + [A_{\hat{a}} , A_{\hat{b}} ] -
C^{\hat{c}}_{\ \hat{a} \hat{b}}A_{\hat{c}}.
\end{eqnarray}
Under an infinitesimal $ G $ gauge transformation
$\lambda=\lambda(x^{\mu},X^{\hat{a}})$ we have
\begin{equation}
\delta A_{\hat{a}} = -[ X_{\hat{a}}, \lambda] +
[\lambda,A_{\hat{a}}]~,
\end{equation}
thus $F_{MN}$ is covariant under {local} $G$ gauge
transformations: $F_{MN}\rightarrow F_{MN}+[\lambda, F_{MN}]$.
This is an infinitesimal abelian $U(1)$ gauge transformation if
$\lambda$ is just an antihermitian function of the coordinates
$x^\mu, X^{\hat{a}}$ while it is an infinitesimal non-abelian
$U(P)$ gauge transformation if $\lambda$ is valued in
${\rm{Lie}}(U(P))$, the Lie algebra of hermitian $P\times P$
matrices. In the following we will always assume
${\rm{Lie}}(U(P))$ elements to commute with the coordinates
$X^{\hat{a}}$. In fuzzy/non-commutative gauge theory and in
Fuzzy-CSDR a fundamental role is played by the covariant
coordinate,
\begin{equation}
\varphi_{\hat{a}} \equiv X_{\hat{a}} + A_{\hat{a}}~.
\end{equation}
This field transforms indeed covariantly under a gauge
transformation, $
\delta(\varphi_{\hat{a}})=[\lambda,\varphi_{\hat{a}}]~. $ In terms
of $\varphi$ the field strength in the non-commutative directions
reads,
\begin{eqnarray}
F_{\mu \hat{a}} &=&
\partial_{\mu}\varphi_{\hat{a}} + [A_{\mu}, \varphi_{\hat{a}}]=
D_{\mu}\varphi_{\hat{a}},\\[.3 em]
F_{\hat{a} \hat{b}} &=& [\varphi_{\hat{a}}, \varphi_{\hat{b}}] -
C^{\hat{c}}_{\ \hat{a} \hat{b}} \varphi_{\hat{c}}~;
\end{eqnarray}
and using these expressions the action reads
\begin{equation}
{\cal A}_{YM}= \int d^{4}x\, Tr\, tr_{G}\,\left( {k\over
4g^{2}}F_{\mu \nu}^{2} + {k\over
2g^{2}}(D_{\mu}\varphi_{\hat{a}})^{2} -
V(\varphi)\right),\label{theYMaction}
\end{equation}
where the potential term $V(\varphi)$ is the $F_{\hat{a} \hat{b}}$
kinetic term (in our conventions $F_{\hat{a} \hat{b}}$ is antihermitian so
that $V(\varphi)$ is hermitian and non-negative)
\begin{eqnarray}\label{pot}
V(\varphi)&=&-{k\over 4g^{2}} Tr\,tr_G \sum_{\hat{a} \hat{b}}
F_{\hat{a} \hat{b}} F_{\hat{a} \hat{b}}
\nonumber \\
%
&=&-{k\over 4g^{2}} Tr\,tr_G \left( [\varphi_{\hat{a}},
\varphi_{\hat{b}}][\varphi^{\hat{a}}, \varphi^{\hat{b}}] -
4C_{\hat{a} \hat{b} \hat{c}} \varphi^{\hat{a}} \varphi^{\hat{b}}
\varphi^{\hat{c}} + 2r^{-2}\varphi^{2} \right).
\end{eqnarray}
The action (\ref{theYMaction}) is naturally interpreted as an
action in four dimensions. The infinitesimal $G$ gauge
transformation with gauge parameter $\lambda(x^{\mu},X^{\hat{a}})$
can indeed be interpreted just as an $M^4$ gauge transformation.
We write
\begin{equation}
\lambda(x^{\mu},X^{\hat{a}})=\lambda^{\alpha}(x^{\mu},X^{\hat{a}}){\cal
T}^{\alpha} =\lambda^{h, \alpha}(x^{\mu})T^{h}{\cal
T}^{\alpha}~,\label{3.33}
\end{equation}
where ${\cal T}^{\alpha}$ are hermitian generators of $U(P)$,
$\lambda^{\alpha}(x^\mu,X^{\hat{a}})$ are $n\times n$
antihermitian matrices
and thus are expressible as $\lambda(x^\mu)^{\alpha , h}T^{h}$,
where $T^{h}$ are antihermitian generators of $U(n)$. The fields
$\lambda(x^{\mu})^{\alpha , h}$, with $h=1,\ldots n^2$, are the
Kaluza-Klein modes of $\lambda(x^{\mu}, X^{\hat{a}})^{\alpha}$. We
now consider on equal footing the indices $h$ and $\alpha$ and
interpret the fields on the r.h.s. of (\ref{3.33}) as one field
valued in the tensor product Lie algebra ${\rm{Lie}}(U(n)) \otimes
{\rm{Lie}}(U(P))$. This Lie algebra is indeed ${\rm{Lie}}(U(nP))$
(the $(nP)^2$ generators $T^{h}{\cal T}^{\alpha}$ being $nP\times nP$
antihermitian matrices that are linear independent).
Similarly we rewrite the gauge field $A_\nu$ as
\begin{equation}
A_\nu(x^{\mu},X^{\hat{a}})=A_{\nu}^{\alpha}(x^{\mu},X^{\hat{a}}){\cal
T}^{\alpha} =A_{\nu}^{h, \alpha}(x^{\mu})T^{h}{\cal T}^{\alpha},
\end{equation}
and interpret it as a ${\rm{Lie}}(U(nP))$ valued gauge field on
$M^4$, and similarly for $\varphi_{\hat{a}}$. Finally $Tr\,
tr_{G}$ is the trace over $U(nP)$ matrices in the fundamental
representation.

Up to now we have just performed a ordinary fuzzy dimensional
reduction. Indeed in the commutative case the expression
(\ref{theYMaction}) corresponds to rewriting the initial
lagrangian on $M^4\times S^2$ using spherical harmonics on $S^2$.
Here the space of functions is finite dimensional and therefore
the infinite tower of modes reduces to the finite sum given by
$Tr$.
\section{Non-trivial Dimensional reduction in the case of Fuzzy
Extra Dimensions} Next we  reduce the number of gauge fields and
scalars in the action (\ref{theYMaction}) by applying the Coset
Space Dimensional Reduction (CSDR) scheme. Since $SU(2)$ acts on
the fuzzy sphere $(SU(2)/U(1))_F$, and more in general  the group
$S$ acts on the fuzzy coset $(S/R)_F$, we can state the CSDR
principle in the same way as in the continuum case, i.e. the
fields in the theory must be invariant under the infinitesimal
$SU(2)$, respectively $S$, action up to an infinitesimal gauge
transformation
\begin{equation}
{\cal L}_{\hat{b}} \phi = \delta_{W_{\hat{b}}}\phi= W_{\hat{b}}\phi,
\end{equation}
\begin{equation}
{\cal L}_{\hat{b}}A = \delta_{W_{\hat{b}}}A=-DW_{\hat{b}},
\label{csdr}
\end{equation}
where $A$ is the one-form gauge potential $A = A_{\mu}dx^{\mu} +
A_{\hat{a}} \theta^{\hat{a}}$, and $W_{\hat{b}}$ depends only on
the coset coordinates $X^{\hat{a}}$ and (like $A_\mu, A_a$) is
antihermitian. We thus write
$W_{\hat{b}}=W_{\hat{b}}^{\alpha}{\cal T}^{\alpha},
\,\alpha=1,2\ldots P^2,$ where ${\cal  T}^i$ are hermitian
generators of $U(P)$ and $(W_b^i)^\dagger=-W_b^i$, here
${}^\dagger$ is hermitian conjugation on the $X^{\hat{a}}$'s.

In terms of the covariant coordinate $\varphi_{\hat{d}}
=X_{\hat{d}} + A_{\hat{d}}$ and of
\begin{equation}
\omega_{\hat{a}} \equiv X_{\hat{a}} - W_{\hat{a}}~,
\end{equation}
the CSDR constraints assume a particularly simple form, namely
\begin{equation}\label{3.19}
[\omega_{\hat{b}}, A_{\mu}] =0,
\end{equation}
\begin{equation}\label{eq7}
C_{\hat{b} \hat{d} \hat{e}} \varphi^{\hat{e}} = [\omega_{\hat{b}},
\varphi_{\hat{d}} ].
\end{equation}
In addition we  have a consistency condition  following from the
relation
$[{\cal{L}}_{\hat{a}},{\cal{L}}_{\hat{b}}]=
C_{\hat{a}\hat{b}}^{~~\hat{c}}{\cal{L}}_{\hat{c}}$:
\begin{equation}\label{3.17}
[ \omega_{\hat{a}} , \omega_{\hat{b}}] = C_{\hat{a} \hat{b}}^{\ \
\hat{c}} \omega_{c},
\end{equation}
where $\omega_{\hat{a}}$ transforms as $
\omega_{\hat{a}}\rightarrow \omega'_{\hat{a}} =
g\omega_{\hat{a}}g^{-1}. $ One proceeds in a similar way for the
spinor fields \cite{Aschieri:2003vy, Aschieri:2004vh}.
\subsection{Solving
the CSDR constraints for the fuzzy sphere}We consider
$(S/R)_{F}=S^2_{F}$, i.e. the fuzzy sphere, and to be definite at
fuzziness level $N-1$ ($N \times N$ matrices). We study here the
basic example where the gauge group is $G=U(1)$. In this case the
$\omega_{\hat{a}}=\omega_{\hat{a}}(X^{\hat{b}})$  appearing in the
consistency condition (\ref{3.17}) are $N \times N$ antihermitian
matrices and therefore can be interpreted as elements of
${\rm{Lie}}(U(N))$. On the other hand the $\omega_{\hat{a}}$
satisfy the commutation relations (\ref{3.17}) of
${\rm{Lie}}(SU(2))$. Therefore in order to satisfy the consistency
condition (\ref{3.17}) we have to embed ${\rm{Lie}}(SU(2))$ in
${\rm{Lie}}(U(N))$. Let $T^h$ with $h = 1, \ldots ,(N)^{2}$ be the
generators of ${\rm{Lie}}(U(N))$ in the fundamental
representation, we can always use the convention $h= (\hat{a} ,
u)$ with $\hat{a} = 1,2,3$ and $u= 4,5,\ldots, N^{2}$ where the
$T^{\hat{a}}$ satisfy the $SU(2)$ Lie algebra,
\begin{equation}
[T^{\hat{a}}, T^{\hat{b}}] = C^{\hat{a} \hat{b}}_{\ \
\hat{c}}T^{\hat{c}}~.
\end{equation}
Then we define an embedding by identifying
\begin{equation}
 \omega_{\hat{a}}= T_{\hat{a}}.
\label{embedding}
\end{equation}
The constraint (\ref{3.19}), $[\omega_{\hat{b}} , A_{\mu}] = 0$,
then implies that the four-dimensional gauge group $K$ is the
centralizer of the image of $SU(2)$ in $U(N)$, i.e. $$
K=C_{U(N)}(SU((2))) = SU(N-2) \times U(1)\times U(1)~, $$  where
the last $U(1)$ is the $U(1)$ of $U(N)\simeq SU(N)\times U(1)$.
The functions $A_{\mu}(x,X)$ are arbitrary functions of $x$ but
the $X$ dependence is such that $A_{\mu}(x,X)$ is ${\rm{Lie}}(K)$
valued instead of ${\rm{Lie}}(U(N))$, i.e. eventually we have a
four-dimensional gauge potential $A_\mu(x)$ with values in
${\rm{Lie}}(K)$. Concerning the constraint (\ref{eq7}), it is
satisfied by choosing
\begin{equation}
\label{soleasy} \varphi_{\hat{a}}=r \varphi(x) \omega_{\hat{a}}~,
\end{equation}
i.e. the unconstrained degrees of freedom correspond to the scalar
field $\varphi(x)$ which is a singlet under the four-dimensional
gauge group $K$.

The choice (\ref{embedding}) defines one of the possible embedding
of ${\rm{Lie}}(SU(2))$ in ${\rm{Lie}}(U(N))$. For example we could
also embed ${\rm{Lie}}(SU(2))$ in ${\rm{Lie}}(U(N))$ using the
irreducible $N$-dimensional rep. of $SU(2)$, i.e. we could
identify $\omega_{\hat{a}}= X_{\hat{a}}$. The constraint
(\ref{3.19}) in this case implies that the four-dimensional gauge
group is $U(1)$ so that $A_\mu(x)$ is $U(1)$ valued. The
constraint (\ref{eq7}) leads again to the scalar singlet
$\varphi(x)$.

In general, we start with a $U(1)$ gauge theory on $M^4\times
S^2_F$. We solve the CSDR constraint (\ref{3.17}) by embedding
$SU(2)$ in $U(N)$. There exist $p_{N}$ embeddings, where $p_N$ is
the number of ways one can partition the integer $N$ into a set of
non-increasing positive integers \cite{Mad}. Then the constraint
(\ref{3.19}) gives the surviving four-dimensional gauge group. The
constraint (\ref{eq7}) gives the surviving four-dimensional
scalars and eq. (\ref{soleasy}) is always a solution but in
general not the only one. By setting
$\phi_{\hat{a}}=\omega_{\hat{a}}$ we obtain always a minimum of
the potential. This minimum is given by the chosen embedding of
$SU(2)$ in $U(N)$.

\section{Discussion and Conclusions}
Non-commutative Geometry has been regarded as a promising
framework for obtaining finite quantum field theories and  for
regularizing quantum field theories. In general quantization of
field theories on non-commutative spaces has turned out to be much
more difficult and with less attractive ultraviolet features than
expected \cite{Filk:dm, Minwalla:1999px}, see however ref.
\cite{Grosse:2004ik}, and ref. \cite{Steinacker}. Recall also
that non-commutativity is not the only suggested tool for
constructing finite field theories. Indeed four-dimensional finite
gauge theories have been constructed in ordinary space-time and
not only those which are ${\cal N} = 4$ and ${\cal N} = 2$
supersymmetric, and most probably phenomenologically
uninteresting, but also chiral ${\cal N} = 1$ gauge theories
\cite{Kapetanakis:vx} which already have been successful in
predicting the top quark mass and have rich phenomenology that
could be tested in future colliders
\cite{Kapetanakis:vx,Kubo:1994bj}. In the present work we have not
adressed the finiteness of non-commutative quantum field theories,
rather we have used non-commutativity to produce, via Fuzzy-CSDR,
new particle models from  particle models on $M^4\times (S/R)_F$.

The Fuzzy-CSDR has different features from the ordinary CSDR
leading  therefore to new four-dimensional particle models.
It may well be that Fuzzy-CSDR provides more
realistic four-dimensional theories. Having in mind the
construction of realistic models one can also combine the fuzzy
and the ordinary CSDR scheme, for example considering $M^4\times
S'/{R'}\times (S/R)_F$.

A major difference between fuzzy and ordinary SCDR is that in the
fuzzy case one always embeds $S$ in the gauge group $G$ instead of
embedding just $R$ in $G$. This is due to the fact that the
differential calculus on the fuzzy coset space is based on $dim S$
derivations instead of the restricted $dim S - dim R$ used in the
ordinary one.  As a result the four-dimensional gauge group $H =
C_G(R)$ appearing in the ordinary CSDR after the geometrical
breaking and before the spontaneous symmetry breaking due to the
four-dimensional Higgs fields does not appear in the Fuzzy-CSDR.
In Fuzzy-CSDR the spontaneous symmetry breaking mechanism takes
already place by solving the Fuzzy-CSDR constraints. The
four-dimensional potential has the typical ``maxican hat'' shape,
but it appears already spontaneously broken. Therefore in four
dimensions appears only the physical Higgs field that survives
after a spontaneous symmetry breaking. Correspondingly in the
Yukawa sector of the theory we have the results of the spontaneous
symmetry breaking, i.e. massive fermions and Yukawa interactions
among fermions and the physical Higgs field. Having massive
fermions in the final theory is a generic feature of CSDR when $S$
is embedded in $G$ \cite{Kapetanakis:hf}. We see that if one would
like to describe the spontaneous symmetry breaking of the SM in
the present framework, then one would be naturally led to large
extra dimensions.

A fundamental difference between the ordinary CSDR and its fuzzy
version is the fact that a non-abelian gauge group $G$ is not
really required in high dimensions. Indeed  the presence of a
$U(1)$ in the higher-dimensional theory is enough to obtain
non-abelian gauge theories in four dimensions.

The final point that we would like to stress here is the question
of the renormalizability of the gauge theory defined on $M_4
\times (S/R)_F$. First we notice that the theory exhibits certain
features so similar to a higher-dimensional gauge theory defined
on $M_4 \times S/R$ that naturally it could be considered as a
higher-dimensional theory too. For instance the isometries of the
spaces $M_4 \times S/R$ and $M_4 \times (S/R)_F$ are the same. It
does not matter if the compact space is fuzzy or not. For example
in the case of the fuzzy sphere, i.e. $M_4 \times S^2_F$, the
isometries are $SO(3,1) \times SO(3)$ as in the case of the
continuous space, $M_4 \times S^2$. Similarly the coupling of a
gauge theory defined on $M_4 \times S/R$ and on $M_4 \times
(S/R)_F$ are both dimensionful and have exactly the same
dimensionality. On the other hand the first theory is clearly
non-renormalizable, while the latter is renormalizable (in the
sense that divergencies can be removed by a finite number of
counterterms). So from this point of view one finds a partial
justification of the old hopes for considering quantum field
theories on non-commutative structures. If this observation can
lead  to finite theories too, it remains as an open question.

\section*{Acknowledgements}
We would like to thank L. Castellani and H. Steinacker for useful
discussions. Two of us, (J.M.) and (G.Z.), would like to thank the
organizers of the III Summer School in Modern Mathematical
Physics, Zlatibor, Serbia, 20-31.08.2004, for the very warm
hospitality.



\end{document}